\documentclass{KapProc}
\normallatexbib

\usepackage{epsfig} 

\begin{document}


\articletitle{Random Matrix Theory of Scattering\\
in Chaotic and Disordered Media}

\author{Jean-Louis Pichard}
\affil{Service de Physique de l'Etat Condens\'e, \\
CEA-Saclay, 91191 Gif sur Yvette cedex, France}
\email{pichard@drecam.saclay.cea.fr}

\vskip1cm
\noindent
{\bf Published in:} {\it Waves and Imaging through Complex Media}\, 
ed.\ by P.~Sebbah, Kluwer Academic Publishers (2001).

\begin{abstract}
We review the random matrix theory describing elastic
scattering through zero-dimensional ballistic cavities (having
chaotic classical dynamics) and quasi-one dimensional disordered
systems. In zero dimension, general symmetry considerations
(flux conservation and time reversal symmetry) are only considered,
while the combination law of scatterers put in series is taken into account
in quasi-one dimension. Originally developed for calculating the
distribution of the electrical conductance of mesoscopic systems,
this theory naturally reveals the universal behaviors
characterizing elastic scattering of various scalar waves.
\end{abstract}



This chapter is a short introductory review of the random matrix
descriptions of elastic scattering. Additional informations can
be found in more exhaustive reviews
\cite{PichardRev,StoneRev,MelloRev}. The more recent reviews are
given by Bohigas \cite{BohigasRev}, by Beenakker \cite{BeenakkerRev} 
and by Guhr, M\"uller-Groeling and Weidenm\"uller \cite{GuhrRev}. 
The basic references for random matrix theory are the book of Mehta
\cite{Mehta} (see also Porter \cite{Porter}) and the series of
papers published by Dyson \cite{Dyson} in 1962.

\section{Gaussian ensembles of Hermitian matrices}
 
 For a statistical description of a matrix ensemble, one has to 
define a measure of the space of the matrices having the required 
symmetries. If one is interested by the distribution of a restricted 
set of parameters suitable for describing the matrices (e. g. the 
eigenvalues of an Hermitian matrix), one has to use the system of 
coordinates using those parameters. The J acobian of the transformation 
(e. g. from matrix elements coordinates towards eigenvalue-eigenvector 
coordinates) yields correlations between those parameters. Those correlations 
(level repulsions) are at the origin of universal behaviors first observed 
in complex nuclei, then in small metallic particles, quantum billiards, 
hydrogen atom in a magnetic field, mesoscopic quantum systems, 
electro-magnetic cavities... 
The simplest illustration is given by the Gaussian ensembles of Hermitian 
matrices introduced in this section. Another illustration is given in 
the following section: the distribution of the radial parameters 
characterizing a scattering matrix $S$ or a transfer matrix $M$.

   For doing statistics with real numbers, one defines the
probability $P(dx)$ to have a real number $x$ inside an
infinitesimal interval of length $dx$: $P(dx)=\rho(x) \mu(dx)$
where $\rho(x)$ is a density and $\mu(dx)=dx$ the measure of an
infinitesimal interval of the real axis. Similarly, for doing
statistics with matrices $X$, one defines the measure $\mu(dX)$
of an infinitesimal volume element $dX$ of the matrix space in which
$X$ is defined and one gives a density probability $\rho(X)$.
The measure $\mu(dX)$ is given by the symmetries of $X$,
while the density $\rho(X)$ may contain physical assumptions
(e.g. minimum information density given a few physical constraints).

For instance, let us introduce the Gaussian Orthogonal Ensemble
(GOE) of real symmetric matrices $H$ and the probability distribution
of their eigenvalues $E_i$. A real symmetric matrix $H=H^T=H^*$ of size
$N$ has $(N^2+N)/2$ independent entries. The infinitesimal volume
element $dH$ has a measure $\mu(dH)$ given by the product of the
infinitesimal variations $dH_{ij}$ of the $N(N+1)/2$ independent entries:
\begin{equation}
\mu(dH)= \prod_{i \leq j}^N dH_{ij}.
\end{equation}
A possible definition of the GOE density probability $\rho(H)$ is
given by a maximum entropy criterion.
Minimizing \cite{Balian} the information entropy
\begin{equation}
I(\rho(H)) = -\int \rho (H) \ln \rho(H) \mu(dH)
\end{equation}
with an imposed expectation value for the trace of $H^2$
gives
\begin{equation}
\rho(H) \propto \exp \left(- \frac{tr H^2}{ 2 \sigma^2}\right).
\end{equation}
Since
$$
\exp \left(- \frac {tr H^2}{2 \sigma^2}\right) \mu (dH) =
\prod_{i=1}^N \exp \left(-\frac{H_{ii}^2}{2 \sigma^2}\right)
dH_{ii} \prod_{i<j}^N \exp \left(-\frac{
H_{ij}^2}{\sigma^2}\right) dH_{ij},
$$
the $N(N+1)/2$ independent matrix elements $H_{ij}$ ($i\leq j$)
are uncorrelated variables with Gaussian distributions of
variance $\sigma^2$ and $\sigma^2/2$ for the diagonal and the
off-diagonal entries respectively. This ratio between the
variances is important since it makes the GOE ensemble invariant
under change of basis: $\rho(H)$ is only function of the $N$
eigenvalues $E_i$ of $H$ through $\sum_{i=1}^{N} E_i^2$ and does
not depend on the eigenvectors of $H$. To calculate $P(E_1,
\ldots, E_N)$, one has to go from the parameterization of $H$ in
terms of its matrix elements $H_{ij}$ to the parameterization of
$H$ in terms of its eigenvalue-eigenvector coordinates. The
Jacobian of this change of coordinates is at the basis of the
level repulsion and of the spectral rigidity characteristic of
usual random matrix theories.

A real symmetric matrix is diagonalizable by an orthogonal 
transformation. Let us define the measure of an orthogonal 
transformation $O_N$. We first consider a $2d$ rotation $O_2$ 
of angle $\theta$. One has
\begin{eqnarray}
O_2 =
\left(
\begin{array}{cc}
\cos \theta & \sin \theta \\
-\sin \theta & \cos \theta
\end{array}
\right)
\end{eqnarray}
and by differentiation
\begin{eqnarray}
d O_2 = O_2
\left(
\begin{array}{cc}
0 & d\theta \\
-d\theta & 0
\end{array}
\right)
\end{eqnarray}

Clearly, $\mu (dO_2)=d \theta$ is the appropriate measure for 
the transformation $O_2$. The generalization to an arbitrary 
$N \times N$ orthogonal transformation $O_N$ is straightforward:
 
\noindent
\( \begin{array} {clcr}
O_N^T O_N= I_N & (O_N + dO_N)^T (O_N+dO_N) = I_N \\
dO_N = O_N dA_N & dA_N^T =- dA_N
\end{array}\)

\noindent
 where $I_N$ denotes the unit $N \times N$ matrix.
$dA_N$ is a $N \times N$ real antisymmetric matrix and
$\mu(dO_N)=\prod_{i<j}^N dA_{ij}$.

$H$ is diagonalizable by an orthogonal transformation $O$: $H=O
H_D O^T$ where $H_D$ is a real diagonal matrix of entries $E_i$
and of measure $\mu(dH_D)= \prod_{i=1}^N dE_i$. By
differentiation, one gets
\begin{eqnarray}
dH= O d{\cal H} O^T  \\
d{\cal H}=dA H_D - H_D dA + dH_D
\end{eqnarray}
where we have used $dO=O dA$ and $dA^T = -dA$. The real symmetric
matrix $dH$ is related to $d{\cal H}$ by an orthogonal
transformation. The Jacobian is equal to one and
$\mu(dH)=\prod_{i \leq j}^N dH_{ij}= \prod_{i \leq j}^N d{\cal
H}_{ij}$. The product of the infinitesimal diagonal elements of
$d{\cal H}$ gives $\mu(dH_D)$, the off-diagonal contribution
gives $ \prod_{i<j}^N |E_i-E_j| dA_{ij}$, and one eventually
obtains the measure $\mu(dH)$ in terms of the measures $\mu(dO)$
and $\mu(dH_D)$
\begin{equation}
\mu(dH)= \prod_{i<j}^N |E_i-E_j| \mu(dH_D) \mu(dO).
\end{equation}
In terms of the eigenvalue-eigenvector coordinates of $H$, the GOE
distribution becomes:
\begin{equation}
P(dH)\mu(dH)=P(E_1, \ldots, E_N) \mu (dH_D) \mu (dO)
\end{equation}
where the joint probability distribution of the eigenvalues is
identical to the Gibbs factor of a set of $N$ point charges free
to move on the real axis of the complex plane with a pairwise
logarithmic repulsion and a quadratic confining potential at an
inverse temperature $\beta=1$:
\begin{equation}
P(E_1, \ldots, E_N) \propto \exp \left[-\beta \sum_{i<j}^N \ln
|E_i-E_j| + \sum_{i=1}^N \frac{E_i^2} {2 \sigma^2}\right]
\end{equation}
The pairwise repulsion coming from the Jacobian makes unlikely 
level degeneracies and explains why dramatically non-random an 
energy-level series really is. 
This is the Coulomb gas analogy usual in Random Matrix Theory. To
appreciate how this random matrix approach is adapted to include
symmetry breaking effects, let us assume that the matrix $H$ is
the Hamiltonian of an electron moving in a chaotic cavity.
Applying a magnetic field removes time reversal symmetry, $H$
becomes hermitian ($H=H^{\dagger}$) and $\mu(dH)=\prod_{i=1}^N
dH_{ii}^1 \prod_{i<j} dH_{ij}^1 dH_{ij}^2$, taking into account
the infinitesimal variations of the real and imaginary parts of
its matrix elements ($H_{ij}=H_{ij}^1 + i H_{ij}^2$). $H$ is now
diagonalizable by a unitary transformation $U$ and $dU=U da$
where $da$ is an infinitesimal anti-hermitian matrix
($da=-da^{\dagger}$), and $\mu(dU)=\prod_{i=1}^N da_{ii}^2
\prod_{i<j}^N da^1_{ij}da^2_{ij}$. One obtains for hermitian
matrices $dH=U d{\cal H} U^{\dagger}$ where $d{\cal H} = da H_D +
dH_D - H_D da$. The Jacobian of a unitary transformation being
equal to one, one eventually finds $\mu(dH) = \prod_{i<j}
|E_i-E_j|^2 \mu(dH_D) \mu(dU) /(\prod_i da_{ii}^2)$, the square
coming from the fact that the non diagonal contribution of
$d{\cal H}$ is now complex. Breaking time reversal symmetry, one
keeps the Coulomb gas analogy, with a temperature divided by a
factor two ($ \beta=1 \rightarrow 2$). For electrons of spin
$1/2$, one can also break spin rotation symmetry (SRS) by spin
orbit scattering, an effect which preserves time reversal
symmetry (TRS). The matrix elements of $H$ are no longer real
($\beta=1$) or complex ($\beta=2$), but quaternion real
$(\beta=4)$ and the level distribution is still given by the
Coulomb gas analogy with a temperature divided by a factor 4:
$\beta=1 \rightarrow 4$.

The main feature of those three Gaussian ensembles of random matrices
is that $\rho(H)$ does not couple eigenvalue and eigenvector variables.
These ensembles are invariant under canonical transformations: orthogonal
transformations ($\beta=1)$ when the system invariant under TRS and SRS
symmetries, unitary transformations $(\beta=2$) in the
absence of TRS and symplectic transformations ($\beta=4$) with TRS and without
SRS. The eigenvectors are totally random, the measure of the matrices
$O$ or $U$ are given by the Haar measures over the orthogonal or unitary
groups respectively, and the integration over the eigenvectors is trivial.
This is the totally random character of the eigenvectors
which makes the energy levels correlated (pairwise logarithmic repulsion)
and subject to universal symmetry breaking effects (e.g. $\beta=1,4
\rightarrow 2$ when TRS is broken).

\section{Radial parameterization of scattering matrices $S$ and Measures}

Similar random matrix theories can be adapted to matrices having
different symmetries and can give the joint probability
distribution of a subset of variables which can be used for their
parameterization. Another example is provided by the unitary
matrices describing complex elastic scatterers. Let us consider a
perfect waveguide characterized by $N$ quantized modes
propagating to the right and $N$ time reversed modes propagating
to the left. Let us introduce in the middle of this wave guide a
complex elastic scatterer described by its $2N \times 2N$
scattering matrix $S$. This matrix describes the various
transmission and reflection amplitudes to the right or to the
left, has to be unitary for conserving the flux amplitudes
($SS^{\dagger}=S^{\dagger}S=I_{2N}$), and must be symmetric
($S=S^T$) if one has time reversal invariant scattering:
\begin{eqnarray}
S=\left(
\begin{array}{cc}
r & t' \\ t & r'
\end{array}
\right)
\end{eqnarray}
The $N \times N$ matrices $t$ and $t'$ describe transmission
amplitudes of the incoming fluxes to the right and left
directions respectively, while $r$ and $r'$ describe reflections.
Let us consider time reversal invariant scattering where $S$ is
symmetric and can be decomposed as
\begin{eqnarray}
S=U^T.U    \\   S+dS = U^T ( I_{2N} + i dM) U
\end{eqnarray}
where $U$ and $dM$ are respectively a unitary and an
infinitesimal real symmetric matrices. This decomposition is not
unique, one can multiply $U$ by an arbitrary orthogonal
transformation $O$ ($U \rightarrow UO$) but the measure
$\mu(dS)=\prod_{i\leq j} dM_{ij}$ is uniquely defined since the
Jacobian of the transformation $dM \rightarrow d{\cal M} =
OdMO^T$ is equal to one. $\mu(dS)$ was expressed by Dyson
\cite{Dyson} in terms of the eigenvectors and eigenvalues of $S$
and their associated measures. The original motivation in Dyson's
work was not at all to study a scattering problem, but to use the
eigenvalue distribution of $S$ for describing energy-level
statistics (the eigenvalues of $S$ being confined on the unit
circle of the complex plane, one does not need the somewhat
artificial GOE quadratic confining potential). For a scattering
problem, the eigenvalue-eigenvector parameterization of $S$ is
not adapted and we introduce a more convenient one using $2$
unitary $N \times N$ matrices $u_1$ and $u_2$ and a diagonal  $N
\times N$ matrix $\Lambda$ with $N$ real positive diagonal entries
$\lambda_1, \ldots, \lambda_N$. Denoting $T=(1+\Lambda)^{-1}$ and
$R=\Lambda(1+\Lambda)^{-1}$, $S$ can be written as:
\begin{eqnarray}
S=
\left(
\begin{array}{cc}
u_1  &  0 \\ 0  & u_2
\end{array}
\right)
\left(
\begin{array}{cc}
-{\sqrt R}  &  {\sqrt T} \\ {\sqrt T}  & {\sqrt R}
\end{array}
\right)
\left(
\begin{array}{cc}
u_1^T  &  0 \\ 0  & u_2^T
\end{array}
\right)
\end{eqnarray}
In this parameterization, the transmission and reflection
matrices become
\begin{eqnarray}
t=u_2 {\sqrt T} u_1^T   &  t'=u_1 {\sqrt T} u_2^T \\
r=-u_1{\sqrt R} u_1^T   &  r'= u_2 {\sqrt R} u_2^T
\end{eqnarray}
and $T$ and $R$ contain the eigenvalues of $tt^{\dagger}$ and
$rr^{\dagger}$ (transmission and reflection eigenvalues). One can
note that the transfer matrix $M$ which gives the flux amplitudes
of the right side of the scatterer in terms of the flux
amplitudes of its left side can be written using the same
parameterization as
\begin{eqnarray}
M=
\left(
\begin{array}{cc}
u_2  &  0 \\ 0  & u_2^*
\end{array}
\right)
\left(
\begin{array}{cc}
{\sqrt {I_N+\Lambda}}  &  {\sqrt \Lambda} \\
{\sqrt \Lambda}  & {\sqrt{I_N+\Lambda}}
\end{array}
\right)
\left(
\begin{array}{cc}
u_1^T  &  0 \\ 0  & u_1^{\dagger}
\end{array}
\right)
\end{eqnarray}
$M$ is pseudo unitary (flux conservation)
\begin{eqnarray}
M
\left(
\begin{array}{cc}
I_N  &  0 \\ 0  & -I_N
\end{array}
\right)
M^{\dagger}
=
M^{\dagger}
\left(
\begin{array}{cc}
I_N  &  0 \\ 0  & -I_N
\end{array}
\right)
M
=
\left(
\begin{array}{cc}
I_N  & 0 \\ 0  & -I_N
\end{array}
\right)
\end{eqnarray}
and when one has TRS ($\beta=1$), $M$ must also satisfy the
requirement:
\begin{eqnarray}
M^*
=
\left(
\begin{array}{cc}
0  &  I_N \\ I_N  & 0
\end{array}
\right)
M
\left(
\begin{array}{cc}
0  &  I_N \\ I_N  & 0
\end{array}
\right).
\end{eqnarray}
$M$ has the advantage to be multiplicative when one puts the
scatterers in series. For the derivation of this parameterization
of $S$, see Ref.\cite{Pichard-Mello}, and use the relation
between the matrices $M$ and $S$. To understand the interest 
of this parameterization, we introduce the variables  $x_i$ 
from $\lambda_i=\sinh^2 x_i$. One can see than $M$ is decomposed 
in the product of a unitary transformation, followed by $N$
hyperbolic rotations of angle $x_i$:
\[
\left(
\begin{array}{cc}
{\sqrt {1+\lambda_i}}  &  {\sqrt \lambda_i} \\
{\sqrt \lambda_i}  & {\sqrt{1+\lambda_i}}
\end{array}
\right)
\rightarrow
\left(
\begin{array}{cc}
\cosh x_i  &  \sinh x_i \\ \sinh x_i  & \cosh x_i
\end{array}
\right)
\]
before a second unitary transformation. 
The $N$ parameters $\lambda_i$ are called the radial parameters
of $M$ (or $S$).

Diagonalizing the matrix containing those parameters in the new
parameterization of $S$ by the orthogonal matrix $O$
\begin{eqnarray}
O=\frac{1}{\sqrt2}
\left(
\begin{array}{cc}
(I_N-\sqrt{R})^{1/2}  &  (I_N+\sqrt{R})^{1/2} \\
(I_N-\sqrt{R})^{1/2}  &  -(I_N+\sqrt{R})^{1/2}
\end{array}
\right)
\end{eqnarray}
and denoting
\begin{eqnarray}
U=
\left(
\begin{array}{cc}
u_1  &  0 \\ 0  & u_2
\end{array}
\right)
\end{eqnarray}
and
\begin{eqnarray}
I=
\left(
\begin{array}{cc}
iI_N  &  0 \\ 0  & -iI_N
\end{array}
\right)
\end{eqnarray}
one can write $S=YY^T$ where the unitary matrix $Y=UOI$. Defining
the infinitesimal anti-hermitian and real antisymmetric matrices
$dA$ and $dB$ from $dU=UdA$ and $dO=OdB$, using $dS=iYdMY^T$, one
obtains:
\begin{eqnarray}
idM & = & dC+dC^T \\
dC & = & I^*(dB+O^TdAO)I
\end{eqnarray}
which allows us to write the Jacobian matrix of the change of
coordinates $dM_{ij} \rightarrow (dA_{ij},dB_{ij})$. This matrix
has a simple block diagonal form and its determinant gives the
measure $\mu(dS)$ in terms of the measures
$\mu(d\Lambda)=\prod_{i=1}^N  d\lambda_i$ and
$\mu(dU)=\prod_{i=1}^2\mu(du_i)$. We have sketched the derivation
when $S$ is unitary symmetric ($\beta=1)$, but the extension to
the three possible symmetry classes is straightforward and gives
for $\mu_{\beta}(dS)$ the general form \cite{JPB-BM}
\begin{eqnarray}
\mu_{\beta}(dS)=P_{\beta} (\lambda_1, \ldots, \lambda_N) \mu(d\Lambda)
\mu(dU)  \\
P_{\beta} (\lambda_1, \ldots, \lambda_N) = \exp-\beta
H_{\beta} (\lambda_1, \ldots, \lambda_N)  \\
H_{\beta} (\lambda_1, \ldots, \lambda_N)=-\sum_{i<j}^N \ln |\lambda_i - \lambda_j| + \sum_{i=1}^N V_{\beta} (\lambda_i) \\
V_{\beta} (\lambda)=(N + \frac{2-\beta}{2\beta}) \ln(1+\lambda)
\end{eqnarray}
If the scattering is not time reversal symmetric ($\beta=2$), $S$
is no longer symmetric and one needs two additional unitary
matrices $u_3$ and $u_4$ for parameterizing $S$ and
$\mu(dU)=\prod_{i=1}^4 \mu(du_i)$. If one considers scattering of
spin $1/2$ particles by a TRS scatterer which removes SRS
(spin-orbit scattering), $\beta=4$.

One can similarly show that the measure for the transfer
matrices \cite{StoneRev}
is given in terms of the radial parameters $\lambda_i$ by:
\begin{eqnarray}
\mu_{\beta}(dM)=\prod_{i<j}^N |\lambda_i-\lambda_j|^{\beta} \mu(d\Lambda)
\prod_{i=1}^{2,4} \mu(du_i)
\end{eqnarray}

\section{Zero dimensional chaotic scattering}

\begin{figure}[h]
\begin{center}
\includegraphics*[height=6cm]{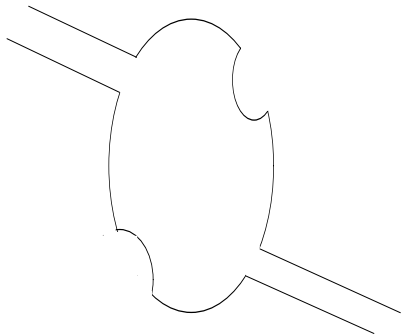}
\end{center}
\caption{Scheme of a cavity giving rise to zero-dimensional 
chaotic scaterring.}
\label{Fig1}
\end{figure}

Let us assume that the scatterer represented by the matrix $S$ is
a ballistic cavity of irregular shape having chaotic classical
dynamics, as sketched in Fig. \ref{Fig1}. 
We have in mind long scattering trajectories corresponding to
particles reflected many times inside the cavity before being
transmitted or reflected. Let us assume that the shape of the cavity
is slightly modified, or that the wave length of the incoming fluxes
varies. The scattering will be deeply re-organized and a statistical
description of the fluctuations of the scattering amplitudes becomes
necessary. To this end, we need to define a statistical ensemble of
scattering matrices $S$ and we will assume that the scatterer will
visit this ensemble when one varies a tunable parameter (shape of
the cavity, wave-length, applied magnetic field reorganizing the
quantum interferences if we consider electron elastic scattering).
The simplest ensemble is the one where all the scattering processes
are equiprobable, which does not contain any information about
the system excepted its basic symmetries. Those ensembles of
minimum information entropy for $S$ are the circular ensembles
\cite{Dyson} introduced by Dyson in 1962, for which the probability
to find $S$ inside a volume element $dS$ is
\begin{equation}
P(dS)=\frac{1}{V_{\beta}} \mu_{\beta} (dS)
\end{equation}
where $V_{\beta}$ is a normalization constant. One obtains for
the radial parameters ${\lambda_i}$ a Coulomb gas analogy very
similar to the GOE-GUE-GSE Coulomb gas analogies for the energy
level of a random Hamiltonian, excepted two noticeable
differences: (i) the $\lambda_i$ are real positive in contrast to 
the $E_i$ which are only real, the Coulomb gas is free to move only 
on the positive part of the real axis in the complex plane (ii) the confining
potential $V_{\beta}(\lambda)$ is implied by the symmetries of
$S$ (in contrast to the quadratic potential given by a certain
choice of $\rho(H)$) and depends on the symmetry parameter
$\beta$.

Let us see the implication for the total transmission probability
${\cal T}=tr tt^{\dagger}=\sum_{i=1}^N T_i=\sum_{i=1}^N (1+\lambda_i)^{-1}$.
When $S$ gives Fermi wave scattering by a mesoscopic scatterer coupling to
electron reservoirs, ${\cal T}$ gives its electrical
conductance in units of $2e^2/h$ (Landauer formula).

\begin{figure}[h]
\begin{center}
\includegraphics*[height=6cm]{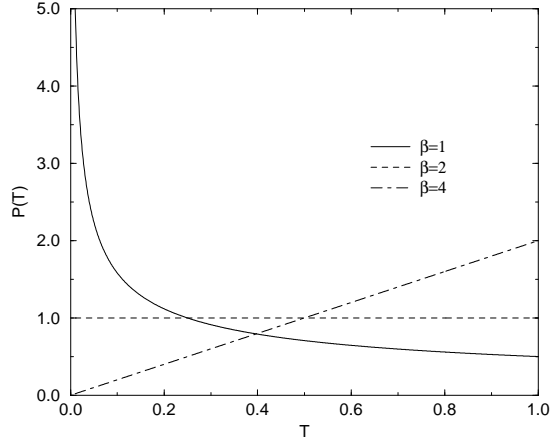}
\end{center}
\caption{Symmetry breaking effect on the distribution $P({\cal T})$ 
for a chaotic cavity coupled to leads via single mode contacts ($N=1$).}
\label{Fig2}
\end{figure}

If $N=1$ (single mode wave guide), the probability distribution of ${\cal T}$
exhibits strong symmetry breaking effects (see Fig. \ref{Fig2}):
\begin{equation}
P({\cal T}) =\frac{\beta}{2} {\cal T}^{\beta/2-1}
\end{equation}

If one measures the electrical conductance of a chaotic cavity
(quantum billiard) coupled to two electron reservoirs by single
mode leads, the conductance is likely to be close to $0$ if one
has TRS and SRS, to have a value uniformly distributed between
$0$ and $2e^2/h$ if one applies a small magnetic field (no TRS),
and a value close to $2e^2/h$ if the electron reflection on the
walls of the cavity is accompanied by spin-orbit scattering (no
SRS). The distribution $P({\cal T})$ can be explored by changing
the Fermi energy with a metallic gate, by small deformations of
the cavity or by applying a magnetic field.

If $N >>1$ the symmetry breaking effects are much smaller. A first
one is the suppression, when there is no TRS, of a small ``weak
localization'' correction to the average transmission $<{\cal
T}>$, the second effect of removing TRS is to halve the universal
variance of ${\cal T}$.

\subsection{Weak Localization Corrections}

The large $N$ limit of the density $\rho(\lambda)$
\begin{equation}
\rho_{\beta}(\lambda)=\sum_{i=1}^N \int_{R^+} \ldots \int_{R^+}
\prod_{i=1}^N d\lambda_i
P_{\beta} (\lambda_1, \ldots, \lambda_N) \delta(\lambda-\lambda_i)
\end{equation}
can be calculated using an equation derived by Dyson \cite{Dyson72}
\begin{equation}
\int_{R^+} \frac{\rho_{\beta}(\lambda') d\lambda'}{\lambda-\lambda'}
+\frac{\beta-2}{2\beta} \frac{d \ln \rho_{\beta}(\lambda)}{d\lambda}
=\frac{dV_{\beta}(\lambda)}{d\lambda}
\label{Dyson}
\end{equation}
Taking the $V_{\beta}(\lambda)$ characterizing the three circular
ensembles for $S$, one gets for the ensemble averaged total
transmission:
\begin{equation}
<{\cal T}>=\int_{R^+} \frac{\rho_{\beta}(\lambda)}{1+\lambda}=\frac{N}{2}
+\frac{\beta-2}{4\beta}+O(\frac{1}{N})
\end{equation}
The term $\propto N$ is obvious: having chaotic scattering, the
probability to be reflected equals the probability to be
transmitted. However there is a small correction of order $1$
which reduces by $-1/4$ ${\cal T}$ when there is TRS and SRS,
which disappears without TRS and which enhances ${\cal T}$ by a
factor $1/8$ without SRS. This is the analog in a quantum
billiard of the well-known weak-localization corrections to the
Boltzmann-Drude conductance of a disordered system.

\subsection{Universal Conductance Fluctuations}

One of the main phenomena which is naturally explained by random matrix
theory \cite{MPS} are the ``universal conductance fluctuations'' (UCF)
characterizing mesoscopic conductors. When one varies the ``conductance''
${\cal T}$ with an external parameter (Fermi wave length, magnetic field ...)
one generates fluctuations of magnitude independent of $<{\cal T}>$. To
calculate $<\delta {\cal T}^2>$ one needs to know the density-density
correlation function of the $\lambda$-parameters. In the limit
$N \rightarrow \infty$, one can simply calculate \cite{beenakker} this
variance. Exploiting the Coulomb gas analogy, one can write
\begin{equation}
K_{\beta}(\lambda,\lambda')=
<\sum_{ij}\delta(\lambda-\lambda_i)\delta(\lambda-\lambda_j)>
-\rho_{\beta}(\lambda)\rho_{\beta}(\lambda')
\end{equation}
as a functional derivative:
\begin{equation}
K_{\beta}(\lambda,\lambda')=-\frac{1}{\beta}
\frac{\partial \rho_{\beta}(\lambda)}{\partial V_{\beta} (\lambda')}
\end{equation}
When $N \rightarrow \infty$, $V_{\beta} (\lambda) \rightarrow
\int_{R^+} \ln |\lambda-\lambda'| \rho_{\beta} (\lambda) + const$
(this amounts to neglect the term responsible for the weak
localization correction in Eq. (\ref{Dyson}). In the large
$N$-limit, $V_{\beta}(\lambda)$ becomes a linear functional of
$\rho(\lambda')$. This implies an important result: $K_{\beta}
(\lambda, \lambda')$ depends only on the nature of the pairwise
repulsion between the radial parameters and becomes independent
of the confining potential $V_{\beta} (\lambda)$. The evaluation
of the functional derivative giving $K_{\beta}(\lambda,\lambda')$
is straightforward:
\begin{equation}
\lim_{N\rightarrow \infty} K_{\beta} (\lambda,\lambda') = -\frac
{1}{\pi^2\beta} \ln \frac {{\sqrt \lambda}-{\sqrt \lambda'}}
{{\sqrt \lambda}+{\sqrt\lambda'}}.
\end{equation}
to eventually give
\begin{equation}
<\delta^2({\cal A})> \rightarrow \frac {-1}{\beta \pi^2}
\int_{R^+}\int_{R+}   d\lambda d\lambda'
\ln \frac {{\sqrt \lambda}-{\sqrt \lambda'}}
{{\sqrt \lambda}+{\sqrt \lambda'}}
\frac{da(\lambda)}{d\lambda}\frac{da(\lambda')}{d\lambda'}
\end{equation}
for the variance $<\delta^2 A>$ of a linear statistics
$A=\sum_{i=1}^N a (\lambda_i)$ of the $\lambda$-parameter. Taking
$a(\lambda)=(1+\lambda)^{-1}$, one gets
\begin{equation}
<\delta^2{\cal T}> = \frac {2}{16 \beta}
\end{equation}
The variance of ${\cal T}$ is just a number which depends on
$\beta$ but not on $<{\cal T}>$.

When $N$ is finite, weak localization corrections and variances
can be exactly calculated using a method introduced by Mehta,
Gaudin and Dyson. Using a set of orthogonal polynomials
($\beta=2$) or shew-orthogonal polynomials ($\beta=1,4$), one can
perform the integration of $P_{\beta} (\lambda_1, \ldots,
\lambda_N)$ over an arbitrary set of $\lambda$-parameters in
order to have $\rho_{\beta}(\lambda)$, $K_{\beta}(\lambda,
\lambda')$ and higher order correlation functions. Going to the
variables $T_i=(1+\lambda_i)$ the polynomials required
\cite{JPB-BM} to perform the integrals for $\beta=2$ are the
Legendre polynomials.

\section{Many channel disordered wave guide}

\begin{figure}[h]
\begin{center}
\includegraphics*[height=6cm]{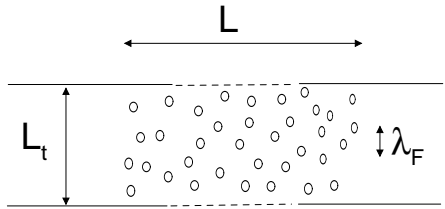}
\end{center}
\caption{Scheme of a quasi-$1d$ disoredered wave guide of length 
$L >> L_t$ with $N=(L_t/\lambda_F)^{d-1}$ modes.}
\label{Fig3}
\end{figure}

So far, we have done the simplest random matrix exercise where
all the scatterers are taken with a uniform distribution. Those
circular ensembles are suitable to describe scattering in zero-dimensional
chaotic cavities, as it has been numerically checked. Another exactly
solvable case is provided by a $1d$ series of scatterers. Let us consider
a quasi-$1d$ disordered wire (or wave guide) with $N$ channels, as 
sketched in Fig. \ref{Fig3}. 

Typically,
if the (Fermi) wavelength is $\lambda_F$, and $L_t^{d-1}$ the transverse
section, one has $N=(L_t/\lambda_F)^{d-1}$. Let us consider
a ``building block'' of length $\delta L$ and of transfer matrix
$M_{\delta L}$. A suitable statistical ensemble for $M_{\delta L}$ is
given by a maximum entropy ensemble where $<\cal T>$ is imposed. If
$\delta L$ is larger than the elastic mean free path $l$, a natural
requirement is to impose Ohm's law (the conductance $<g>\propto <{\cal T}>
\propto Nl(\delta L)^{-1}$). This gives for the building block
$P (M_{\delta L}) \propto \exp -(A <{\cal T}>) \mu(dM)$, where $A$
is the Lagrange multiplier associated to the imposed constraint. This
ensemble preserves the logarithmic pairwise interaction between
the $\lambda$-parameter and the distribution of the auxiliary unitary
matrices $u_i$ derived for the circular ensembles, but gives rise
to a different potential $V_{\beta}(\lambda)$. Considering a $1d$ series
of such building blocks, of length $L/l$, and exploiting the multiplicative
combination law of $M$ ($M_{L+\delta L} = M_{L}.M_{\delta L}$), one
can derive a Fokker-Planck equation \cite{BeenakkerRev,Dorokhov,MPK} for
$P_{\beta} (\lambda_1, \ldots, \lambda_N,L/l)$:
\begin{eqnarray}
\frac {\partial P_{\beta} (\lambda_1, \ldots, \lambda_N)}{ \partial L} =
D \Delta_{\lambda}  P_{\beta} (\lambda_1, \ldots, \lambda_N) \\
D  =  \frac {2}{(\beta N +2 -\beta)l} \\
\Delta_{\lambda} = \sum_{i=1}^N {\frac{\partial}{\partial \lambda_i}}
\lambda_i (1+\lambda_i) J {\frac {\partial}{\partial \lambda_i}}
{\frac {1}{J}} \\
J  =  \prod_{i<j}^N |\lambda_i-\lambda_j|^{\beta}
\end{eqnarray}
In this statistical description of $M$ (or $S$), the radial
parameters and the matrices $u_i$ are statistically
uncorrelated.  The $u_i$ remain distributed with Haar measure
over the $N \times N$ unitary group. Increasing the number $L/l$
of blocks put in series only changes the statistics of the radial
parameters as given by the above Fokker-Planck equation, This is
the limitation of this ``isotropic'' model which allows us to
solve it entirely and to describe quasi-$1d$ localization. But
the transverse system dimensions appear only through the
parameter $N$. A strip of purely $1d$ transverse section and a
bar with a $2d$ section are treated the same way: within the $0d$
approximation for the transverse dynamics. When $N=1$, $J=1$ and
the resulting equation was originally derived
\cite{gertsenshtein-vasilev} in 1959 in a work entitled:
``waveguides with random inhomogeneities and Brownian motion in
the Lobachevsky plane''. The nature of the Fokker-Planck equation
was indeed correctly identified as a ``Heat equation in a space of
negative curvature''. Looking in the mathematical literature, one
can realize that $\Delta_{\lambda}$ is the radial part of the
Laplace-Beltrami operator in the space of the transfer matrices
$M$. Increasing $L$ yields for $M$ an ``isotropic''  (invariant
under the unitary transformation $u_i$) Brownian motion in the
transfer matrix space which is only characterized by the
diffusion constant $D$.

\subsection{Quasi-$1d$ Localization}

When $L$ increases and exceeds the localization length $\xi$, the
system becomes an Anderson insulator and it is more convenient to use
the variables $x_i=L/\xi_i$ ($\lambda_i = \sinh^2 x_i$), where the
lengths $\xi_i$ characterize the exponential decays of the transmission
channels of the quasi-$1d$ scatterer ($T_i \approx
\exp -(2L/\xi_i)$). When $L \rightarrow \infty$ the variables
$1/\xi_i$ are given by the Lyapounov exponents of the multiplicative
transfer matrix $M$, giving $N$ decay lengths $\xi_N << \xi_{N-1} << \xi_1$,
the largest of them defining \cite{pichard-sarma} the localization length
$\xi$ of the quasi-$1d$ system. The Fokker-Planck equation becomes:
\begin{eqnarray}
\frac {\partial P_{\beta} (x_1, \ldots, x_N)}{ \partial L} =
D \sum_{i=1}^N {\frac{\partial}{\partial x_i}} \left( P_{\beta}
+\beta P_{\beta} {\frac {\partial \Omega}{\partial x_i}}\right) \\
\Omega  =  \sum_{i<j}^N \ln |\sinh x_i^2 - \sinh^2 x_j| -\frac{1}{\beta}
\sum_{i=1}^N \ln |\sinh x_i|
\end{eqnarray}
where the different channels are coupled via $\Omega$. However,
when $L \rightarrow \infty $, $|\sinh^2 x_i - \sinh^2 x_j|
\approx \exp 2 x_i $ when $x_i >> x_j$, the channels become
decoupled and $\Omega \rightarrow -(2/\beta) \sum_{i=1}^N
(1+\beta N -\beta) x_i$. The Fokker-Planck equation becomes
solvable and gives \cite{PichardRev}:
\begin{eqnarray}
P_{\beta} (x_1, \ldots, x_N, L) = ({\frac {\gamma l}{2 \pi L}})^{\frac{N}{2}}
\prod_{i=1}^N \exp \left(-{\frac{\gamma l}{2L}} (x_i-{\frac{L}{\xi_i}})^2
\right)
\end{eqnarray}
where $\xi_i = (\gamma l)/ (1+\beta N-\beta)$ and $\gamma = \beta
N +2 -\beta$. One gets two important results:

(i) a universal symmetry breaking effect for the localization length $\xi$
when $\beta=1 \rightarrow 2$:
\begin{equation}
\xi = (\beta N +2 - \beta)l
\end{equation}

(ii) a normal distribution for $\ln {\cal T}$ where $-2 <\ln {\cal
T}> = <\delta^2 \ln {\cal T}>$.
Those symmetry breaking effects
have been observed in magneto-transport measurements performed in
disordered wires \cite{PSSD,PMS,gershenson}, where the
conductance $g \propto \exp -(2L/\xi)$ and $\xi =(N+1)l
\rightarrow 2Nl$ when $\beta=1 \rightarrow 2$. However this
theory neglecting electron-electron interactions, it would be
important to check those universal symmetry breaking effects in
the localized regime using other waves (light, sound ...).

\subsection{Mapping onto a Calogero-Sutherland model of interacting fermions}

When $L \leq \xi$, the system is a disordered conductor where
a number $N_{eff} < N$ of channels are still opened, giving
$<{\cal T }> \approx Nl/L$ (Ohm's law) if one ignores the (small)
weak-localization corrections. Approximations can be used also in
this limit ($L << \xi$) to calculate the quasi-$1d$ weak-localization
corrections $\delta {\cal T} = (\beta-2) / (3\beta)$ and the UCF variance
$2/(15 \beta)$. One can notice that those values valid for the
quasi-$1d$ wire are close, but not identical to those derived from the
circular ensembles. The small difference between the UCF variances
tells us (see the arguments given for having $K (\lambda, \lambda')$ in the
large $N$-limit) that the radial parameters cannot have \cite{beenakker}
exactly the pairwise logarithmic repulsion in quasi-$1d$. This can be
understood when $\beta=2$
where one can solve the Fokker-Planck equation for any values of $N$, using a
transformation originally introduced by Sutherland to solve Dyson's Brownian
motion model. The distribution $P_{\beta} (x_1, \ldots, x_N, L)$ is related
\cite{BeenakkerRev,Beenakker-Rejaei} to a wave function
$\Psi(x_1, \ldots, x_N, L)$ by the transformation
\begin{equation}
P=\Psi \exp -\left ({\frac {\beta \Omega}{2}} \right)
\end{equation}
and the Fokker-Planck equation for $P$ becomes a Schr\"odinger equation
for $\Psi$ in imaginary time.

\begin{equation}
-l \frac {\partial \Psi}{ \partial L} = H \Psi
\end{equation}

where the Hamiltonian is given by

\begin{eqnarray}
H= {\frac {-1}{2\gamma}} \sum_{i=1}^N \left ({\frac{\partial^2}
{\partial x_i^2}} +{\frac{1}{\sinh^2 2x_i}} \right) + (\beta-2) U(x_i,x_j)
\\
U(x_i,x_j)= \frac {\beta}{2\gamma} \sum_{i<j}^N
{\frac {\sinh^2 2x_j + \sinh^2 2x_i}{(\cosh 2x_j-\cosh^2 2x_i)^2}}
\end{eqnarray}

For $\beta=2$, the ``particles'' do not interact and the equation
can be solved to give \cite{Beenakker-Rejaei} a small change in
the Coulomb gas analogy for the radial parameters:
$$
-\ln |\lambda_i-\lambda_j| \rightarrow -\frac {1}{2} \ln
|\lambda_i- \lambda_j| - \frac {1}{2} \ln | \rm{arsinh}^2
\sqrt{\lambda_i} - \rm{arsinh}^2 \sqrt {\lambda_j}|
$$
yielding the change $<\delta^2 {\cal T}> = \frac{2}{16 \beta} \rightarrow
\frac{2}{15 \beta}$.

For $\beta \neq 2$, one has not yet found how to solve the
Schr\"odinger equation. Let us note that the Calogero-Sutherland
Hamiltonian derived from the original Dyson's Brownian motion
model is now solved \cite{Pasquier} when $\beta/2=p/q$ where $p$
and $q$ are integer.

\section{Summary}

We have seen how to derive the probability distributions of $S$
and how to extract the distribution of ${\cal T}$ using very little
information (symmetries, combination law in quasi$1d$). This can
be simply done as far as the radial parameters are decoupled from the
auxiliary matrices $u_i$, which limits the method to $0d$ and quasi-$1d$.
A more difficult task remaining to be achieved is to go beyond this limit,
and to describe elastic scattering in $2d$ and $3d$, including possible
Anderson localization. One can ask to what extend a real chaotic cavity
or a quasi-$1d$ disordered wave-guide is accurately described by those
random matrix theories. Numerical calculations \cite{BM} of the distribution
of ${\cal T}$ in (suitably) designed chaotic cavities (suitably)
connected to two many-channel ballistic waveguides confirm the random matrix
results. For quasi-$1d$ disordered wires, a field theory approach has been
derived \cite{efetov-larkin} assuming local diffusive dynamics. One obtains
a non-linear $\sigma$-model which gives using supersymmetry $<{\cal T (L)}>$
and $<\delta ^2 {\cal T} (L)>$ in the large $N$-limit. The results turn out
to coincide \cite{Frahm,FP} with those given by the Fokker-Planck equation
in this limit.
The finite $N$ behaviors which are calculable by the Fokker-Planck equation
are still out of reach using the $\sigma$-model, as well as the moments of
$\ln {\cal T}$ which are the statistically meaningful observables
(related to a normal distribution) in the localized limit.

\begin{acknowledgments}
My own research on this random matrix theory adapted to scattering
was published in a series of papers done with C.W.J. Beenakker,
Y. Imry, R. A. Jalabert, P. A. Mello, K. Muttalib, K. M. Frahm,
K. Slevin, A. D. Stone and N. Zanon,
collaborations which are gratefully acknowledged.
\end{acknowledgments}

\end{document}